**Title:**

Electrical detection of magnetic skyrmions by non-collinear magnetoresistance

**Authors:**

Christian Hanneken[1], Fabian Otte[2], André Kubetzka[1], Bertrand Dupé[2], Niklas Romming[1], Kirsten von Bergmann[1], Roland Wiesendanger[1] & Stefan Heinze[2]

**Affiliations:**

[1]Department of Physics, University of Hamburg, Jungiusstrasse 11, 20355 Hamburg, Germany

[2]Institute of Theoretical Physics and Astrophysics, Christian-Albrechts-Universität zu Kiel, Leibnizstrasse 15, 24098 Kiel, Germany

**Abstract:**

Magnetic skyrmions are localised non-collinear spin textures with high potential for future spintronic applications[1-12]. Skyrmion phases have been discovered in a number of materials[9,11] and a focus of current research is the preparation, detection, and manipulation of individual skyrmions for an implementation in devices[6-8]. Local experimental characterization of skyrmions has been performed by, e.g., Lorentz microscopy[3] or atomic-scale tunnel magnetoresistance measurements using spin-polarised scanning tunneling microscopy[7,12]. Here, we report on a drastic change of the differential tunnel conductance for magnetic skyrmions arising from their non-collinearity: mixing between the spin channels locally alters the electronic structure, making a skyrmion electronically distinct from its ferromagnetic environment. We propose this non-collinear magnetoresistance (NCMR) as a reliable all-electrical detection scheme for skyrmions with an easy implementation into device architectures.

**Text:**

In magnetic skyrmions, the magnetisation in the centre is opposite to the surrounding ferromagnetic (FM) background (Fig. 1a). Skyrmions are stabilized against collapse by the Dzyaloshinskii-Moriya (DM) interaction[13-15], which arises due to spin-orbit interaction in systems with broken inversion symmetry and imposes a unique rotational sense on the spin structure. Such a particle-like magnetic skyrmion is distinct from the FM state due to its non-trivial topology, and its topological charge can be used as bit of information[5,6]. For the read-out of information encoded in magnetic states various magnetoresistive effects are exploited in current technical devices and sensors. While the giant and the tunnel magnetoresistance (GMR and TMR)[16-19] occur when two magnetic layers are involved, the (tunnel) anisotropic magnetoresistance (T)AMR[20-22] originates from the intrinsic material properties due to spin-orbit coupling. In transport measurements through non-collinear structures, e.g. domain walls, the total resistance deviates from the FM case[23-26]. However, due to the averaging nature of these measurements, discrimination between different contributing effects is difficult[25,26]. In particular, it is unclear to what extent the non-collinearity itself is a source of magnetoresistance.

Our model system is the PdFe atomic bilayer on Ir(111)[7,10,12], which has been studied previously using spin-polarised scanning tunneling microscopy (STM). This method[27] resembles the TMR in STM geometry, where a magnetic tip is separated from the magnetic sample by an insulating vacuum barrier. Figure 1b shows an overview of a PdFe/Ir(111) sample in the skyrmion phase ($B$ = +1.8 T) measured with an STM tip which is not spin-polarised, i.e., the TMR contribution vanishes. Nevertheless, the skyrmions are imaged in the differential tunnel conductance map (d$I$/d$U$, color-coded) as dark circular entities in the PdFe islands. The enlarged d$I$/d$U$ image of Fig. 1c shows two skyrmions at $B$ = -2.5 T and the line profile across one of them (inset) shows that the d$I$/d$U$ signal changes gradually from the FM region to about half of its value at the skyrmion centre. The question of the physical origin of the signal change arises, since neither TMR is present as an unpolarised tip is used, nor is a TAMR contribution expected, because the centre of the skyrmion is collinear with the spins in the FM background (see also Extended Data Fig. 1).

The energy-resolved d$I$/d$U$ signal measured by scanning tunnel spectroscopy can be interpreted as the sample's local density of states (LDOS) in the vacuum[28], which is correlated with the electronic band structure. Figure 1d demonstrates that the local electronic properties of the skyrmion centre deviate significantly from those of the FM background: while the spectrum of the FM state has a peak at about +0.7 V, the centre of the skyrmion exhibits two peaks at about +0.5 V and +0.9 V. We propose that this difference originates from the non-collinearity of the spin structure in the skyrmion.

To validate this hypothesis we perform an experiment in which we change the degree of non-collinearity in a controlled fashion by varying the external magnetic field. The size and shape of a skyrmion in PdFe has been studied as a function of applied field[12] and the polar angle $\theta(d)$ of the magnetisation within a skyrmion is plotted in Fig. 2a as a function of the lateral distance $d$ from its centre for several magnetic field values. We relate the degree of non-collinearity in the centre of a skyrmion with the angle $\alpha$ between a central atom and its neighbouring spins, and find that $\alpha$ scales linearly with $B$, see inset. Figure 2b displays spectra taken at the centre of one skyrmion at different applied fields as indicated, together with reference spectra of the FM background. One can clearly see a systematic shift of the higher energy peak with the applied field. The peak shift $\Delta E$ with respect to the peak of the FM state is roughly linear with $\alpha$ (inset), corroborating our proposal of an effect of the local magnetic non-collinearity on the electronic properties. The laterally resolved d$I$/d$U$ maps at the FM peak energy in Fig. 2c show how the maximum of non-collinearity moves from the rim of the skyrmion to its centre with increasing magnetic field, in agreement with the skyrmion profiles in Fig. 2a.

For the FM state, the experimental d$I$/d$U$ spectra (Fig. 1d and 2b) are in nice agreement with the vacuum LDOS calculated by density functional theory (DFT)[10], see Fig. 3a. The vacuum LDOS is typically dominated by states close to the $\bar{\Gamma}$ point. A detailed analysis of the spin-resolved LDOS and band structure (Extended Data Figs. 2 and 3) reveals that the sharp peak at about +0.8 eV stems from minority $d$-states whereas the rather featureless LDOS of the majority spin channel is due to bands of $s$- and $p$-character.

In a non-collinear spin structure, there is a mixing between the two spin channels resulting in a change of the band structure and the LDOS[29]. This is seen in DFT calculations for the spin spiral phase (Extended Data Figs. 2 and 3, Supplementary Note 2), which are in agreement with the corresponding experimental data (Extended Data Fig. 4, Supplementary Note 3). To capture the key physics of this band mixing for two-dimensional localised skyrmions and to include the skyrmion profiles[12] (cf. Fig. 2a) we use a tight-binding (TB) model. The corresponding Hamiltonian at every atom site is given by

$$H_0 = \begin{pmatrix} \epsilon_\uparrow & 0 \\ 0 & \epsilon_\downarrow \end{pmatrix}, \quad (1)$$

where $\epsilon_\uparrow, \epsilon_\downarrow$ are the on-site energies of the two states. Based on DFT for the FM state we describe the electronic states of PdFe/Ir(111) which dominate the vacuum LDOS using a majority band with a hopping parameter $t_\uparrow = -0.5$ eV, and a minority band with $t_\downarrow = +0.09$ eV and $\epsilon_\uparrow - \epsilon_\downarrow = 3.1$ eV, as depicted in green and red in Fig. 3b. The corresponding spin-resolved LDOS in the vacuum for the FM state is qualitatively very similar to the one obtained by DFT calculations[10], compare Fig. 3c and 3a, and a similar agreement is obtained for spin spiral states (Extended Data Fig. 2). The non-collinearity within the skyrmion leads to a mixing between the majority and the minority spin channels and the hopping between adjacent atomic sites can be described by the matrix

$$V(\alpha) = \begin{pmatrix} t_\uparrow \cos(\alpha/2) & -t_{\uparrow\downarrow} \sin(\alpha/2) \\ t_{\downarrow\uparrow} \sin(\alpha/2) & t_\downarrow \cos(\alpha/2) \end{pmatrix}, \quad (2)$$

where $\alpha$ is the angle between the spins on neighbouring sites and $t_{\uparrow\downarrow} = -t_{\downarrow\uparrow}$ describes the nearest-neighbour hopping matrix element between the two states.

Before solving this TB model for a realistic skyrmion profile, it is instructive to study the effect of the spin mixing in a simplified way. We assume that the matrix $V(\alpha)$ is the same for all atom sites by fixing $\alpha$ and thus obtain a periodic system with a well-defined band structure. Fig. 3b shows that a non-zero value of $\alpha$ (orange line) leads to the formation of a gap in the band structure near the $\bar{\Gamma}$ point where formerly bands were crossing. This leads to the emerging two-peak structure in the vacuum LDOS, see Fig. 3c, in good agreement with the experimental d$I$/d$U$ spectra taken at the skyrmion centre (Fig. 1d).

The energy splitting between the peaks in the vacuum LDOS increases with the angle $\alpha$ between adjacent spins, as demonstrated in Fig. 3d, and $\alpha$ can be correlated to the magnetic-field dependent nearest-neighbour angle in the centre of a skyrmion (cf. Fig. 2a). We include the experimentally determined magnetisation profile of a skyrmion[12] (cf. Fig. 2a) by now choosing $\alpha$ in the matrix $V(\alpha)$ differently for all nearest-neighbour sites of the hexagonal lattice according to the local spin orientation and solve this full nearest-neighbour TB model numerically (cf. methods). Qualitatively, the LDOS at the centre of the skyrmion behaves as in the periodic TB model. However, taking the whole spin structure of the skyrmion into account leads to a larger shift of the high-energy peak with $\alpha$ as shown in Fig. 3e, in better quantitative agreement with the experimental data.

Another means to study the effect of non-collinearity is to use the spatial resolution capabilities of STM to investigate the electronic properties within one skyrmion as a function of distance to the centre. As anticipated from the plot in Fig. 2a, the spectra of a skyrmion at B = -2.5 T show a continuous variation of the peak position from the FM spectrum to the spectrum in the centre of the skyrmion (bottom and top spectrum in Fig. 4a, respectively). The laterally resolved vacuum LDOS from the full TB model (Fig. 4b) also shows a peak shift, in very good agreement with the experiment. To further analyse the effect of non-collinearity on the electronic properties for PdFe/Ir(111), we extract the energy shift $\Delta E$ of the high-energy peak with respect to the FM spectrum for both the experimental data and the calculations and plot it against the lateral distance to the skyrmion centre (Fig. 4c). While the evolution of $\Delta E$ across a skyrmion at $B$ = -2.5 T has a maximum at the centre, the maximum $\Delta E$ for a skyrmion at $B$ = -1 T (Fig. 4d) is much smaller and lies on a circle around the skyrmion centre with a radius of about 1.5 – 2 nm, in agreement with Fig. 2a and 2c.

While in a local STM measurement we obtain a signal change due to NCMR of up to 100% for the small skyrmions at -2.5 T (inset to Fig. 1c), the NCMR will be reduced for spatially averaging planar tunnel junctions. In such a geometry large angles in the centre of a skyrmion are not required or even advantageous, instead one can profit from an increased area of non-collinear spin arrangements. In our system for instance the NCMR of a skyrmion in a 100 nm$^2$ sized junction would be twice as large at -1 T compared to -2.5 T, since the smaller local signal is overcompensated by the larger skyrmion size (compare Fig. 4c,d).

Our work demonstrates the impact of the degree of non-collinearity of a spin structure on the corresponding differential tunnel conductance using unpolarised electrons. As it originates from the spin mixing of bands of opposite spin channels, i.e., a change of the electronic structure, we anticipate that this NCMR effect occurs in a wide range of magnetic materials and is not limited to the tunnel regime. In particular, we propose to use NCMR for the detection of nano-scale skyrmions with a non-magnetic electrode in race-track type spintronic devices with planar tunnel junctions as stationary read heads.

**Methods:**

We apply the Greens function technique[30] to numerically solve the two-band nearest-neighbour TB model given by Eqs. (1) and (2). We use the hopping parameters given in the text and solve the TB model for a two-dimensional hexagonal lattice of (30×30) atom sites with periodic boundary conditions in the $y$-direction and semi-infinite ferromagnetic lattices in the $x$-direction, which is the close-packed direction. The local spin quantization axis on the atomic lattice which determines the angle $\alpha$ for nearest-neighbour sites in Eq. (2) is chosen according to the skyrmion profile[12] $\theta(d)$, see Fig. 2a. We account for the broadening of bands due to the hybridization with the metal substrate by adding an imaginary term $-i\gamma$ to the diagonal elements of our Hamiltonian with $\gamma = 0.1$ eV. The value of $t_{\downarrow\uparrow} = 0.17$ eV was adapted to the experimentally observed peak shift. We have obtained the vacuum LDOS by introducing additional sites in the vacuum. Hopping parameters taking the exponential decay into account connect the atomic lattice sites and the vacuum sites. The hopping matrix element between adjacent vacuum sites leads to a free electron-like dispersion and the on-site energy models the vacuum barrier.

In the periodic TB model, the vacuum decay of the electronic states is taken into account by the exponential factor $\exp(-2z\sqrt{\frac{2m\varphi}{\hbar^2} + \boldsymbol{k}_\parallel^2})$, where $z$ is the distance from the surface, $\varphi$ is the work function, and $\boldsymbol{k}_\parallel$ is the Bloch vector[28]. Due to the $\boldsymbol{k}_\parallel$-dependence, the vacuum decay favors states in the vicinity of the $\bar{\Gamma}$-point of the two-dimensional Brillouin zone as seen in Fig. 3.

**References:**

[1] Bogdanov, A. N. & Yablonskii, D. A. Thermodynamically stable 'vortices' in magnetically ordered crystals. The mixed state of magnets. *Sov. Phys. JETP* **68**, 101-103 (1989).

[2] Mühlbauer, S. *et al.* Skyrmion lattice in a chiral magnet. *Science* **323**, 915-919 (2009).

[3] Yu, X. Z. *et al.* Y. Real space observation of a two-dimensional skyrmion crystal. *Nature* **465**, 901-904 (2010).

[4] Heinze, S. *et al.* Spontaneous atomic-scale magnetic skyrmion lattice in two dimensions. *Nature Phys.* **7**, 713 (2011).

[5] Kiselev, N.S., Bogdanov, A. N., Schäfer, R. & Rößler, U. K. Chiral skyrmions in thin magnetic films: new objects for magnetic storage technology? *J. Phys. D* **44**, 392001 (2011).

[6] Fert, A., Cros, V., & Sampaio, J. Skyrmions on the track. *Nature Nanotech*. **8**, 152-156 (2013).

[7] Romming, N. *et al.* Writing and deleting single skyrmions. *Science* **341**, 636-639 (2013).

[8] Sampaio, J., Cros, V., Rohart, S., Thiaville, A & Fert, A. Nucleation, stability and current-induced motion of isolated magnetic skyrmions in nanostructures. *Nature Nanotech*. **8**, 839-844 (2013).

[9] Nagaosa, N., & Tokura, Y., Topological properties and dynamics of magnetic skyrmions. *Nature Nanotech.* **8**, 899-911 (2013).

[10] Dupé, B., Hoffman, M., Paillard, C. & Heinze, S. Tailoring magnetic skyrmions in ultra-thin transition metal films. *Nat. Commun*. **5**, 4030 (2014).

[11] von Bergmann, K., Kubetzka, A., Pietzsch, O. & Wiesendanger, R. Interface-induced chiral domain walls, spin spirals and skyrmions revealed by spin-polarized scanning tunneling microscopy. *J. Phys.: Condens. Matter* **26** 394002 (2014).

[12] Romming, N., Kubetzka, A., Hanneken, Ch., von Bergmann, K. & Wiesendanger, R. Field-dependent size and shape of single magnetic skyrmions. *Phys. Rev. Lett.* (in press). Preprint available under http://arxiv.org/abs/1504.01573

[13] Dzyaloshinskii, I. E. Thermodynamic theory of "weak" ferromagnetism in antiferromagnetic substances. *Sov. Phys. JETP* **5**, 1259–1262 (1957).

[14] Moriya, T. Anisotropic superexchange interaction and weak ferromagnetism. *Phys. Rev*. **120**, 91–98 (1960).

[15] Fert, A. & Levy, P. A. Role of anisotropic exchange interactions in determining the properties of spin glasses. *Phys. Rev. Lett*. **44**, 1538-1541 (1980).

[16] Baibich, M. N. *et al.* Giant magnetoresistance of (001)Fe/(001)Cr magnetic superlattices. *Phys. Rev. Lett*. **61**, 2472 (1988).

[17] Binasch, G., Grünberg, P., Saurenbach, F. & Zinn, W. Enhanced magneto-resistance in layered magnetic structures with antiferromagnetic interlayer exchange. *Phys. Rev. B* **39**, 4828 (1989).

[18] Jullière, M. Tunneling between ferromagnetic films. *Phys. Lett. A* **54**, 225 (1975).

[19] Wiesendanger, R., Güntherodt, H.-J., Güntherodt, G., Gambino, R.J. & Ruf, R. Observation of vacuum tunneling of spin-polarized electrons with the scanning tunneling microscope. *Phys. Rev. Lett*. **65**, 247 (1990).

[20] McGuire, T.R. & Potter, R.I., Anisotropic magnetoresistance in ferromagnetic 3d alloys. *IEEE Trans. Magn.* **11**, 1018 (1975).

[21] Bode, M. *et al.* Magnetization-direction dependent local electronic structure probed by scanning tunneling spectroscopy. *Phys. Rev. Lett.* **89**, 237205 (2002).

[22] Gould, C. *et al.* Tunneling anisotropic magnetoresistance: a spin-valve-like tunnel magnetoresistance using a single magnetic layer. *Phys. Rev. Lett.* **93**, 117203 (2004).

[23] Levy, P.M. & Zhang, S., Resistivity due to domain wall scattering. *Phys. Rev. Lett*. **79**, 5110 (1997).

[24] Kent, A.D., Yu, J., Rüdiger, U. & Parkin, S.S.P. Domain wall resistivity in epitaxial thin film microstructures. *J. Phys.: Condens. Matter* **13**, 461 (2001).

[25] Marrows, C.H. & Dalton B.C., Spin mixing and spin-current asymmetry measured by domain wall magnetoresistance. *Phys. Rev. Lett*. **92**, 97206 (2004).

[26] Seemann, K.M. *et al.* Disentangling the physical contributions to electrical resistance in magnetic domain walls: a multiscale study. *Phys. Rev. Lett*. **108**, 77201 (2012).

[27] Wiesendanger, R. Spin mapping at the nanoscale and atomic scale. *Rev. Mod. Phys*. **81**, 1495 (2009).

[28] Tersoff, J. & Hamann, D., Theory of the scanning tunneling microscope. *Phys. Rev. B* **31**, 805-813 (1985).

[29] Sandratskii, L. M., Noncollinear magnetism in itinerant-electron systems: theory and applications. *Adv. in Phys.* **47**, 91 (1998).

[30] Datta, S., *Electronic Transport in Mesoscopic Systems*. Cambridge University Press, Cambridge (1995).

**Supplementary Information**

Supplementary Information is available in the online version of the paper.
Supplementary Note 1: Contrast mechanisms in simulated STM images of magnetic skyrmions
Supplementary Note 2: DFT and TB calculations of LDOS for spin spiral states
Supplementary Note 3: NCMR in STM measurements of the spin spiral

**Acknowledgements:**

C.H, A.K, N.R., K.v.B., R.W., S.H. and B.D. acknowledge financial support from the Deutsche Forschungsgemeinschaft via GrK 1286, SFB 668 and project DU1489/2-1. S.H. and B.D. thank the HLRN for providing computational resources. We thank Phivos Mavropoulos, Yuriy Mokrousov, Gustav Bihlmayer and André Kobs for discussions.

**Author contributions:**

C.H. performed the measurements, C.H., K.v.B. and A.K. analysed the experimental data, C.H. and F.O. prepared the figures, K.v.B., A.K. and S.H. wrote the manuscript, A.K. performed the STM simulations. B.D. performed the DFT calculations, F.O. and S.H. devised the TB model, F.O. performed the TB calculations, F.O., B.D. and S.H. analysed the calculations. All authors discussed the results and contributed to the manuscript.

**Author information:**

Reprints and permissions information is available at www.nature.com/reprints. The authors declare no competing financial interests. Correspondence and requests for materials should be addressed to kbergman@physnet.uni-hamburg.de and kubetzka@physnet.uni-hamburg.de.

**Figures (print version and online-only version)**

**Figure 1:**

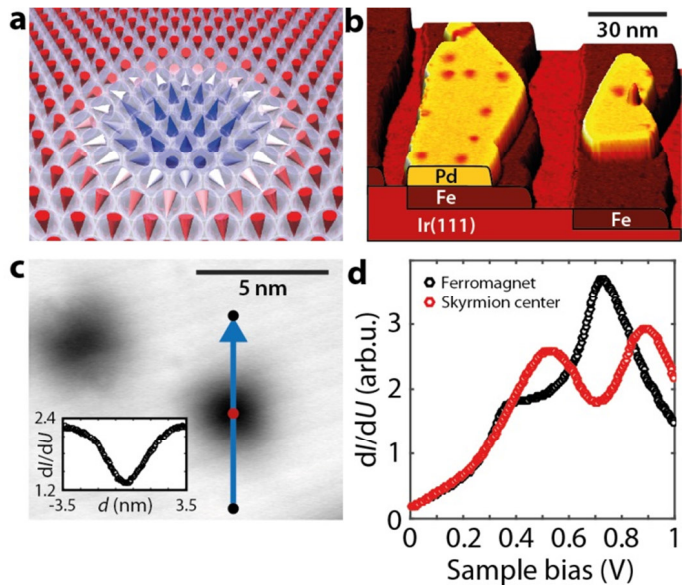

**Figure 1 | Individual skyrmions in PdFe/Ir(111). a,** Sketch of a magnetic skyrmion; cones represent the magnetisation direction. **b,** Perspective view of an STM constant-current image, color-coded with d$I$/d$U$ signal; yellow areas indicate PdFe and red circular entities are magnetic skyrmions ($B$ = +1.8 T, $U$ = +0.7 V, $I$ = 1 nA, $T$ = 8 K). **c,** Closer view of two skyrmions, d$I$/d$U$ map, the inset presents a profile along the arrow ($B$ = -2.5 T, $U$ = +0.7 V, $I$ = 1 nA, $T$ = 4 K). **d,** d$I$/d$U$ tunnel spectra in the centre of a skyrmion (red) and outside the skyrmion in the FM background (black); ($B$ = -2.5 T, $T$ = 4 K, stabilisation parameters $U$ = -1 V, $I$ = 1 nA).

**Figure 2:**

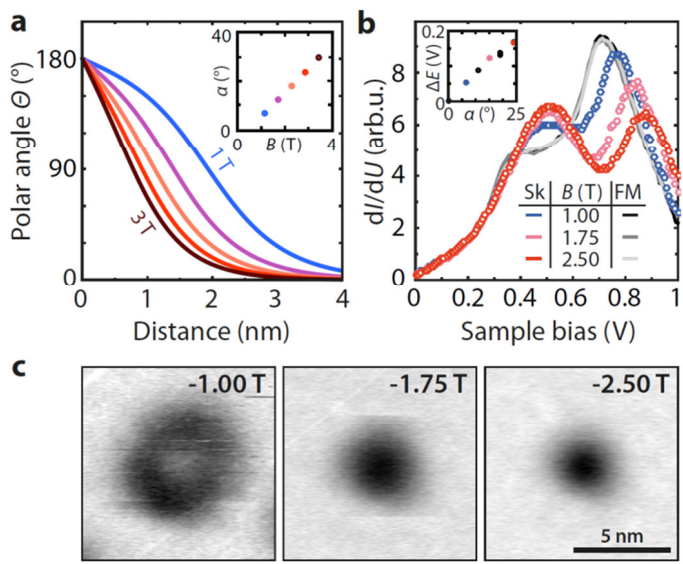

**Figure 2 | Magnetic field-dependent properties of an individual skyrmion. a,** Skyrmion profiles for different magnetic field values, plotted as polar angle $\theta$ of the magnetisation versus distance from the skyrmion centre (based on Ref. 12). **b,** d$I$/d$U$ tunnel spectra measured with a W tip in the centre (Sk) and outside (FM) of an individual skyrmion at different magnetic field values ($T$ = 8 K, stabilisation parameters $U$ = -0.3 V, $I$ = 0.2 nA). **c,** Laterally resolved d$I$/d$U$ maps ($U$ = +0.7 V, $I$ = 1 nA, $T$ = 8 K).

**Figure 3:**

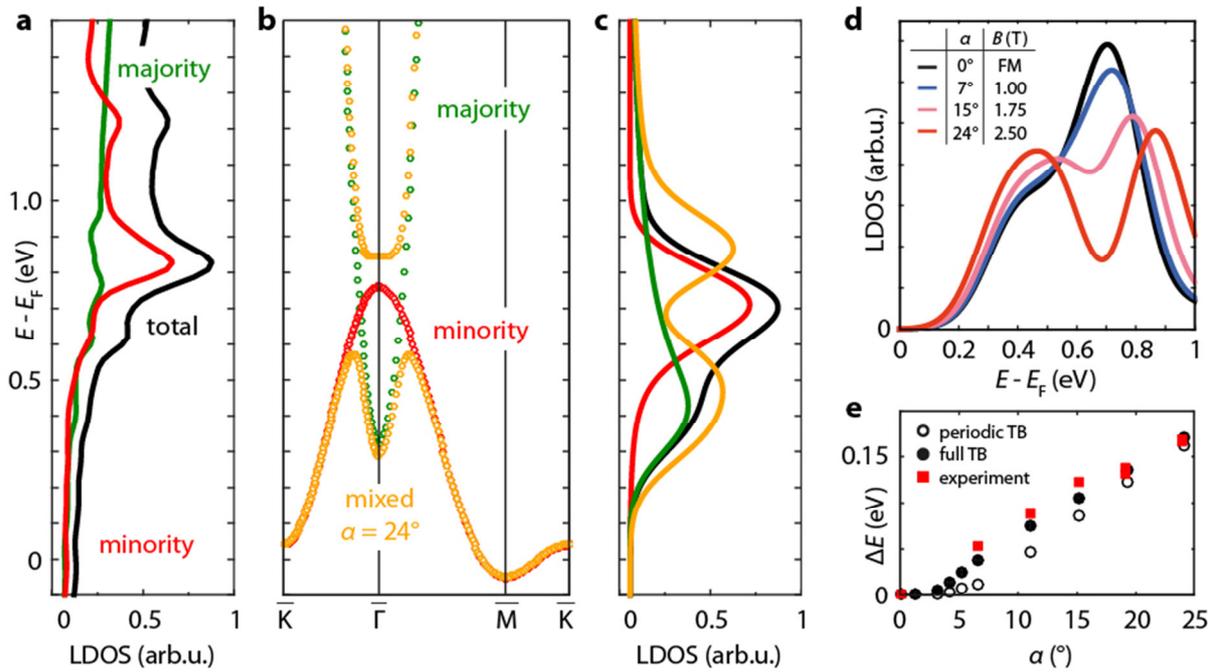

**Figure 3 | Band structure and vacuum LDOS from DFT and TB calculations. a,** Vacuum LDOS calculated from DFT for the FM state. The black line is the total vacuum LDOS and the green and red lines represent the majority and the minority spin channels, respectively. **b,** Band structure from the periodic TB model of the majority (green) and minority (red) spin for the FM state and for $\alpha = 24°$ between nearest neighbour spins (orange, see text for details). **c,** Vacuum LDOS at a distance of 5 Å from the atoms. **d,** TB vacuum LDOS from the periodic TB model at different angles $\alpha$; the data can be correlated to a magnetic field, at which the angle between neighbouring magnetic moments is realised in the centre of a skyrmion, cf. Fig. 2. **e,** Peak shift of the higher energy peak with respect to the FM state for the periodic and the full TB calculations in the centre of the skyrmion, and the experimental data from Fig. 2b for comparison.

**Figure 4:**

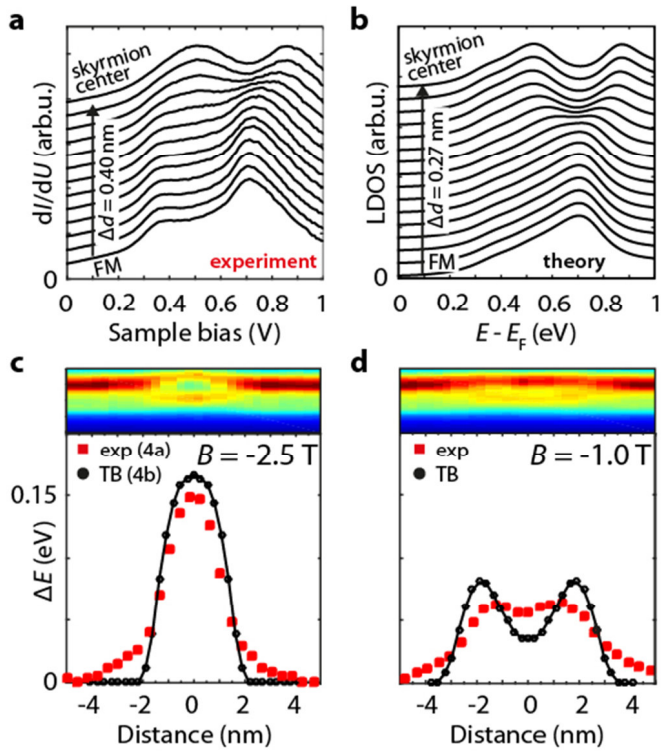

**Figure 4 | Spatial variation of the differential tunnel conductance and calculated vacuum LDOS within a skyrmion. a,** Experimental d$I$/d$U$ tunnel spectra measured with a W tip at different lateral positions $d$ of the skyrmion at a field of $B$ = -2.5 T; the FM spectrum is presented at the bottom and spectra towards the skyrmion centre (top) have a vertical offset for clarity. **b,** Corresponding plot of the vacuum LDOS calculated from the full TB model. **c,** Shift of the high-energy peak as a function of distance to the skyrmion centre extracted from the experimental data (red rectangles) and the full TB model (black circles); the inset at the top shows a colour-coded plot of the data of (a) with energy (vertical) and lateral (horizontal) resolution; red and yellow colour indicate the peak positions. **d,** Same as (c) for $B$ = -1 T. The maximum of $\Delta E$ is now located off-centre at $d$ = 1.5-2.0 nm.

**Extended Data Figures (online-only version):**

**Extended Data Figure 1:**

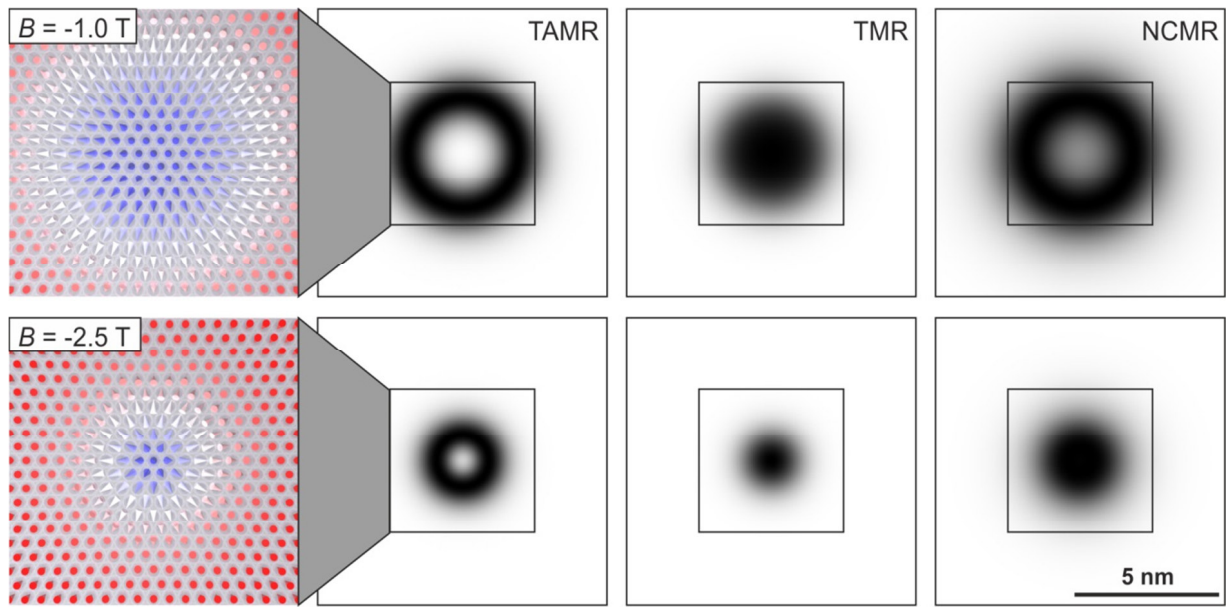

**Extended Data Figure 1 | STM simulations of skyrmions.** Left panels: Spin structure of a skyrmion at two different magnetic field strengths as indicated, each cone represents the atomic spin direction, the colour indicates the out-of-plane component. Other panels: STM simulations for different contrasts as indicated. TAMR contribution to the tunnel current, the in-plane component is chosen to have a lower signal as compared to the out-of-plane component. TMR (SP-STM) simulations for an out-of-plane magnetised tip, where the signal is chosen to be smaller for up-magnetisation components of the sample. NCMR simulations within a simplified model, where the contrast scales with an effective local non-collinearity $\bar{\alpha} = \frac{1}{6}\sum_j \alpha_j$. While at low magnetic fields the NCMR contrast could be mistaken for a TAMR signal, and at high magnetic fields it is qualitatively similar to an SP-STM (TMR) image, the evolution of the NCMR signal with field demonstrates that it is distinct from the two previously known effects (cf. Fig. 2c, details in Supplementary Note 1).

**Extended Data Figure 2:**

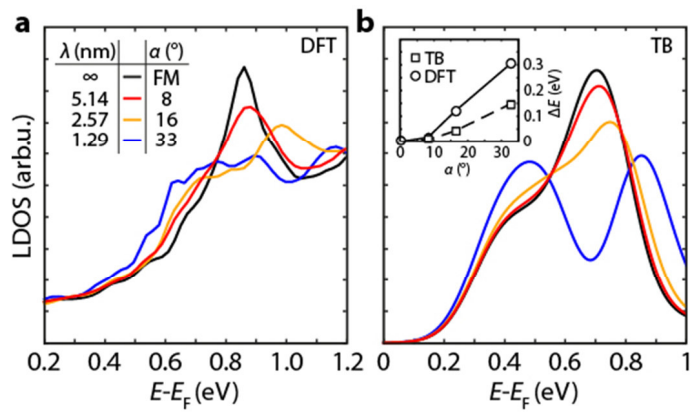

**Extended Data Figure 2 | Vacuum LDOS of spin spiral states. a**, Vacuum LDOS 5 Å above PdFe/Ir(111) above the Fermi energy calculated based on DFT for the FM state ($\lambda = \infty$) and spin spirals with periods of $\lambda = 5.14$ nm, 2.57 nm, and 1.29 nm. **b**, Corresponding vacuum LDOS calculated for spin spirals within the TB model. The inset shows the shift of the higher energy peak obtained from the DFT and from the TB model (details in Supplementary Note 2).

**Extended Data Figure 3:**

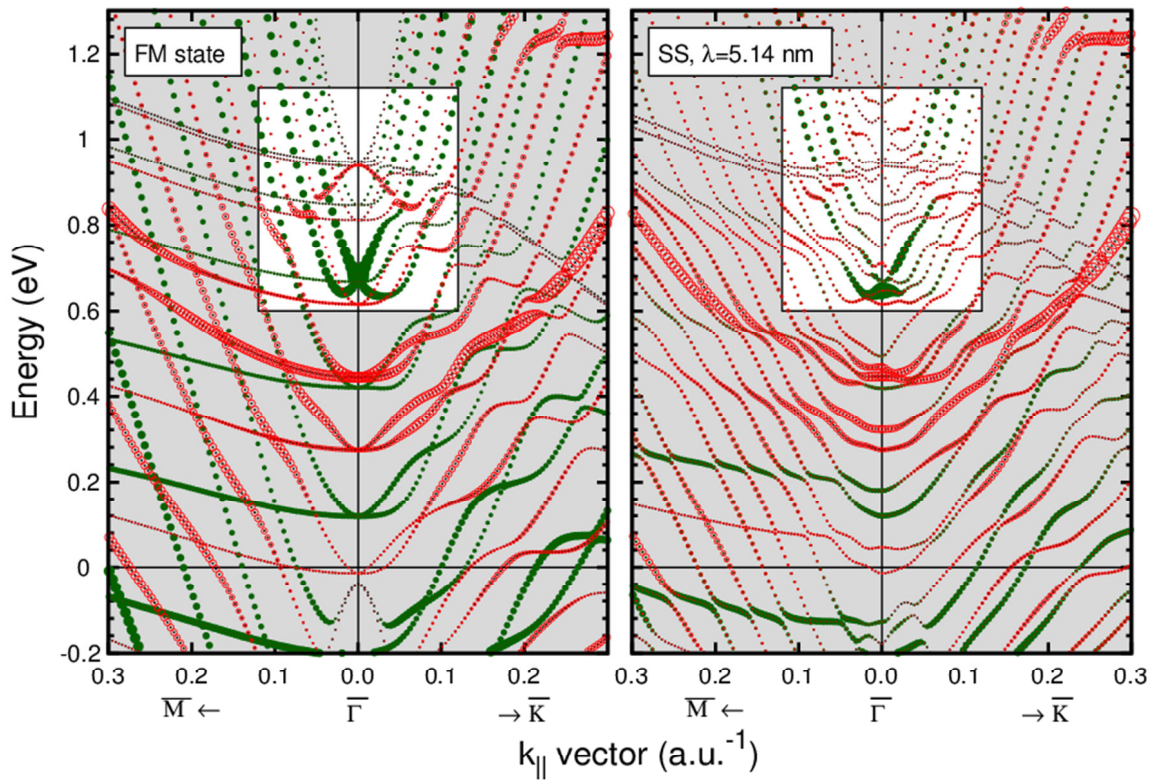

**Extended Data Figure 3 | Band structure of PdFe/Ir(111).** DFT band structure for the FM state and a spin spiral with a period of $\lambda = 5.14$ nm. Bands of spin-up and spin-down character are marked by green and red dots, respectively. The boxes highlight the regime of interest for the peaks in the

vacuum LDOS of Fig. 3a. The band structure is displayed in a larger energy range in order to show that the band mixing does not only occur for states that have a high contribution to the vacuum LDOS (details in Supplementary Note 2).

**Extended Data Figure 4:**

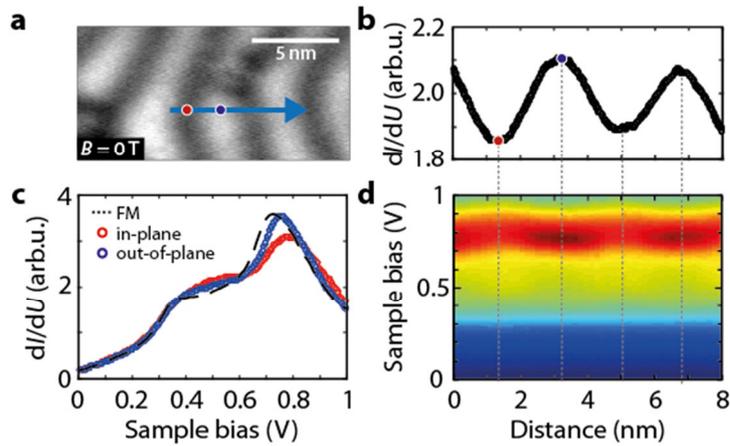

**Extended Data Figure 4 | Spin spiral in PdFe/Ir(111). a,** Map of differential conductance ($B$ = 0 T, $U$ = +0.7 V, $I$ = 1 nA, $T$ = 4 K, Cr bulk tip found to be unpolarised, same tip as used in Fig. 1c,d). **b,** Profile along the blue arrow shown in (a), the period corresponds to half of the magnetic period. **c,** Tunnel spectra at a maximal (blue, out-of-plane magnetisation) and minimal (red, in-plane magnetisation) d$I$/d$U$ signal of the spin spiral, in comparison to the FM spectrum taken at $B$ = -2.5 T (black) ($T$ = 4 K, stabilisation parameters $U$ = -1 V, $I$ = 1 nA). **d,** Spatially and energy resolved d$I$/d$U$ signal (details in Supplementary Note 3).